\RequirePackage[2020-02-02]{latexrelease}
\documentclass[aps,prb]{revtex4}
\usepackage{amsmath,amssymb}
\usepackage{graphicx}
\usepackage{dcolumn}
\usepackage{bm}
\usepackage{color}
\usepackage{relsize}
\newcommand{\mbb}{\mathbb}
\newcommand{\mc}{\mathcal}

\newcommand{\tet}{\texttt}
\newcommand{\pr}{\partial}

\begin{document}
\title{Dynamical optical conductivity for gapped $\alpha-\mc{T}_3$ materials with a curved "flat" band}
\author{
Andrii Iurov$^{1}$\footnote{E-mail contact: aiurov@mec.cuny.edu, theorist.physics@gmail.com
},
Liubov Zhemchuzhna$^{1,2}$,
Godfrey Gumbs$^{2,3}$, 
and
Danhong Huang$^{4}$
}

\affiliation{
$^{1}$Department of Physics and Computer Science, Medgar Evers College of City University of New York, Brooklyn, NY 11225, USA\\ 
$^{2}$Department of Physics and Astronomy, Hunter College of the City University of New York, 695 Park Avenue, New York, New York 10065, USA\\ 
$^{3}$Donostia International Physics Center (DIPC), P de Manuel Lardizabal, 4, 20018 San Sebastian, Basque Country, Spain\\ 
$^{4}$Space Vehicles Directorate, US Air Force Research Laboratory, Kirtland Air Force Base, New Mexico 87117, USA\\ 
}

\date{\today}

\begin{abstract}
We have calculated the dynamical optical conductivity for $\alpha-\mc{T}_3$ materials  in the presence of a finite bandgap in their energy bandstructure. This  is a special type of energy dispersions because for all $\alpha-\mc{T}_3$ materials with a bandgap, except graphene and a dice lattice limits, the flat band receives a non-zero dispersion and assumes a curved shape. The infinite ${\bf k}$-degeneracy of the flat energy band is also lifted. Such a low-energy bandstructure could be obtained if an $\alpha-\mc{T}_3$ material is irradiated off-resonant with circularly polarized light. We have calculated the optical conductivity for the zero and finite temperatures, as well as for the cases of a finite and nearly-zero doping. We have demonstrated that analytical expressions could be in principle obtained for all types of gapped $\alpha-\mc{T}_3$ materials and provided the closed-form analytical expressions for a gapped dice lattice. Our numerical results reveal some well-known signatures of the optical conductivity in $\alpha-\mc{T}_3$ and silicene with two non-equivalent bandgaps, as well as demonstrate some very specific features which have not been previously found in any existing Dirac materials. 
\end{abstract}

\maketitle

\section{Introduction} 
\label{sec1}

$\alpha-\mc{T}_3$ model \,\cite{tabert2015electronic,illes2017properties,bercioux2009massless} describes the electronic behavior of a  wide class of recently discovered two-dimensional lattices with good precision.\,\cite{leykam2018artificial} The atomic build up of an $\alpha-\mc{T}_3$ material consists of a regular hexagon similar to that of graphene (rim atom) with an additional atom at the center of the hexagon which is referred to as a hub atom.\,\cite{raoux2014dia}  Therefore, such  lattices are made up of three atoms per unit cell which results in additional rim-hub electronic hopping and a novel electron low-energy bandstructure.\,\cite{wang2011nearly}

\par
The most unusual feature of this bandstructure is the presence of a flat band in their energy dispersions in addition to a regular Dirac cone which remains stable in the presence of various external factors, such as electric and magnetic fields or impurities.\,\cite{gorbar2019electron} This completely unusual yet relatively simple electronic energy dispersions incited a huge avalanche of research on the electronic,\,\cite{PhysRevB.105.245414,biswas2018dynamics,li2022novel,filusch2021valley,zhou2021andreev,islam2017valley} magnetic,\,\cite{balassis2020magnetoplasmons,illes2016magnetic} optical\,\cite{dey2018photoinduced,iurov2019peculiar,illes2015hall,bryenton2018optical} transport,\,\cite{iurov2022optically,wang2020recovered,kovacs2017frequency} tunneling \,\cite{illes2017klein,iurov2020klein,ye2020quantum,cunha2022tunneling,korol2021chiral,mandhour2020klein} and collective properties \,\cite{malcolm2016frequency,iurov2020many} of $\alpha-\mc{T}_3$ and even $\alpha-\mc{T}_3$-based nanoribbons.\,\cite{tan2020valley,iurov2021tailoring,hao2022zigzag,chen2019enhanced}  with the key question: how are these properties distinguished from the previously studied graphene?\,\cite{gusynin2006magneto,gumbs2014revealing,amado2012magneto,iurov2017effects} 

\par
$\alpha-\mc{T}_3$ model is basically an interpolation between a honeycomb lattice in graphene and a dice lattice with the key parameter $\alpha$ which is a relative hub-rim hopping coefficient or a phase $\phi = \tan^{-1} \alpha$ directly related to the Berry phase of the lattice. Thus, $\alpha = 0$ corresponds to graphene, while $\alpha = 1$ to a dice lattice. 

\medskip
\par 
The most well-known examples of existing materials which completely or partially resemble the $\alpha-\mc{T}_3$ model in their electronic bandstructure are Leib and Kagome optical lattices,\,\cite{xue2019acoustic,mukherjee2015observation,li2018realization,goldman2011topological,julku2016geometric} optical waveguides,\,\cite{constant2016all,yao2014efficient} a trilayer SrTiO$_3$/SrIrO$_3$/SrTiO$_3$,\,\cite{wang2011nearly} Hg$_{1-x}$Cd$_{x}$ quantum well,\,\cite{orlita2014observation} Josephson-junction arrays\,\cite{abilio1999magnetic} and In$_{0.53}$Ga$_{0.47}$As/InP semiconducting layers.\,\cite{franchina2020engineering} A particularly informative and complete review of the fabricated flat band materials was provided in Ref.~[\onlinecite{leykam2018artificial}].

\medskip 
\par 

The bandstructure of $\alpha-\mc{T}_3$ materials becomes even more unusual in the presence of a finite bandgap.\,\cite{cunha2021band} The two subbands corresponding to the valence and conduction bands receive non- equivalent gaps and, most importantly, the flat band assumes a curved shape and is no longer dispersionless. The “flat band” now lies completely below the zero-energy level and has a maximum energy at ${\bf k} = 0$. The only two materials in which the flat band remains actually flat represent the two limiting cases of $\alpha = 0$ and $\alpha = 1$, corresponding to graphene and a dice lattice. Such a non-symmetrical bandgap which affects all the three bands of $\alpha-\mc{T}_3$, may appear due to an external off-resonance dressing field with circular or elliptical polarization.\,\cite{kibis2010metal,kristinsson2016control,iurov2019peculiar}

\medskip
\par 

The conductivity, or, widely speaking, quantum transport is one of the most important and well-investigated properties of an interacting many-body electronic system. If the current does not depend on the variations of the impurities’ position, Boltzmann’s transport theory and the corresponding equation are employed which has been done for all recent two-dimensional materials. The temperature dependence of the Boltzmann conductivity was also investigated for graphene and other lattices.\,\cite{hwang2009screening,iurov2020quantum} The most general approach to conductivity is based on the Kubo formula and on the linear response theory. The calculation normally consists of finding the single particle Green’s function and the current-current (or velocity-velocity) correlation function.

\par
\medskip 

Dynamical, or frequency-dependent optical conductivity is one of the principal electronic properties which was thoroughly investigated for all novel low-dimensional materials: graphene,\,\cite{ziegler2007minimal,stauber2008optical,pellegrino2010strain} silicene,\,\cite{stille2012optical,iurov2018temperature} $\alpha-\mc{T}_3$ and a dice lattice,\,\cite{illes2015hall,kovacs2017frequency,oriekhov2022optical} Kekule-patterned graphene,\,\cite{herrera2020electronic}, transition metal dichalcogenides, \,\cite{tan2021anisotropic,gibertini2014spin,PhysRevB.86.205425,PhysRevB.88.115205} 8-pmmn borophene \,\cite{verma2017effect,napitu2020photoinduced} and twisted bilayers\,\cite{tabert2013optical}, including the situations with magnetic field \,\cite{tabert2013valley, tabert2013magneto,gusynin2006magneto,biswas2016magnetotransport,biswas2018dynamics} Yet, an important and very interesting case of irradiated $\alpha-\mc{T}_3$ with the curved middle band has not been addressed which is the main subject of the present paper. 

\par
The real part of optical conductivity is related to the absorption of photons. When a two-dimensional lattice is irradiated, the transmitted light could be absorbed, exciting the electron from its original energy level. Therefore, calculating and investigating the frequency dependence of the optical conductivity reveals valuable information of the electronic bandstructure of a new material, as well as the structure of all the possible electron transitions between the states which also depends on the doping. A low-energy intraband transitions through the Fermi level is related to a well-known Drude conductivity.\cite{tabert2015electronic}

\medskip
\medskip 
\par
The rest of this paper is structured in the following way. In Sec.~\ref{sec2}, we review and discuss the results for the bandstructure of $\alpha-\mc{T}_3$ in the presence of a $\phi$-dependent energy gap(s), as well as the corresponding electronic states (wavefunctions). Sec.~\ref{sec3} represents different techniques of calculating the optical conductivity for graphene and $\alpha-\mc{T}_3$, which includes finding the electron Green's function and spectral function. Here, we also present and discuss our numerical results for the optical conductivity for the zero and finite temperatures. Our summary, concluding remarks and research perspectives on the subject are provided in Sec.~\ref{sec4}. The detailed calculations and derivations for the single electron states and their energy dispersions in the presence of the finite bandgap, the Green's functions and spectral functions and the optical conductivity in a gapped dice lattice and an arbitrary $\alpha-\mc{T}_3$ material are provided in our appendices \ref{apa}, \ref{apb} and \ref{apc}, correspondingly.

\section{Electronic states with a irradiation-modified bandstructure} 
\label{sec2}

The optical conductivity, which is the main subject of the present paper, mainly depends on the electron bandstructure or, specifically, on the 
on the diverse construct of the multitude of all possible electron transitions from an occupied to an unoccupied electronic state in a specific
material. Therefore, calculating and the analysis of the non-trivial low-energy dispersions of an gapped $\alpha-\mc{T}_3$ lattice is the logical starting point for our investigation.

\medskip

We begin with an overview of the energy dispersions and the corresponding eigenfunctions for $\alpha-\mc{T}_3$ model which are obtained from a pseudospin-1 $\phi$-dependent Dirac-Weyl Hamiltonian model

\begin{equation}
\label{mainH}
\hat{\mc{H}}_\tau^{\phi}(\mbox{\boldmath$k$}) = \hbar v_F \left\{
\begin{array}{ccc}
0 & k^\tau_-\cos \phi & 0 \\
k^\tau_+\cos \phi & 0 & \,  k^\tau_-\sin \phi   \\
0 & k^\tau_+\sin \phi  & 0
\end{array}
\right\}\ ,
\end{equation}
where $v_F = 10^6\,m/s$ is the Fermi velocity equal to that in graphene in order to satisfy $\alpha \longrightarrow 0$ limit of $\alpha-\mc{T}_3$ model,
$\tau = \pm 1$ is the valley index, ${\bf k}=(k_x,k_y)$ is a wave-vector so that $k^\tau_\pm = \tau k_x \pm i k_y$. Hamiltonian \eqref{mainH} 
could be presented as $\hat{\mc{H}}_\tau^{\phi}(\mbox{\boldmath$k$})  = \hbar v_F \, \hat{\mbox{\boldmath$\Sigma$}}(\phi) \cdot {\bf k}$
in terms of $3\times 3$ $\phi-$dependent matrices $\hat{\Sigma}^{(3)}(\phi)$ which we listed and discussed in Appendix \ref{apa}. 

\medskip
Hamiltonian Eq.\,(\ref{mainH}) yields three energy bands $\varepsilon_\gamma^{\,(\phi)}(\mbox{\boldmath$k$}) = \gamma\,\hbar v_F k$, 
corresponding to valence ($ \gamma = - 1$), conduction ($\gamma = + 1$) and flat ($\gamma=0$) bands. These energy bands are clearly independent on
parameters $\tau$ and $\phi$. 

\medskip
The corresponding electronic eigenstates (wavefunctions) are 

\begin{equation}
\label{Eig1}
\Psi^{\gamma=\pm 1}_0(\mbox{\boldmath$k$}\vert\,\tau,\phi)  = \frac{1}{\sqrt{2}} \left\{
\begin{array}{c}
\tau \cos \phi \,\, \tet{e}^{- i \tau \theta_{ \bf k}}  \\
\gamma \\
\tau \sin \phi \,\, \tet{e}^{i \tau \theta_{ \bf k}} 
\end{array}
\right\} \, ,
\end{equation}
where $\theta_{\bf k} = \arctan (k_y/k_x)$.

\begin{equation}
\label{Eig2}
\Psi^{\gamma=0}_0(\mbox{\boldmath$k$}\vert\,\tau,\phi) = \left\{
\begin{array}{c}
\tau\sin \phi \,\, \tet{e}^{- i \tau \theta_{\bf k}}  \\
0 \\
- \tau\cos \phi \,\, \tet{e}^{i \tau \theta_{\bf k}} 
\end{array}
\right\} \, .
\end{equation}
The energy dispersion of dressed-state quasiparticles directly depend on the Berry phase, or phase $\phi$, and valley index $\tau$ in contrast to 
energy dispersions $E^{\,(0)}_{\gamma}(\mbox{\boldmath$k$})$.

\medskip
\par
\begin{figure} 
\centering
\includegraphics[width=0.6\textwidth]{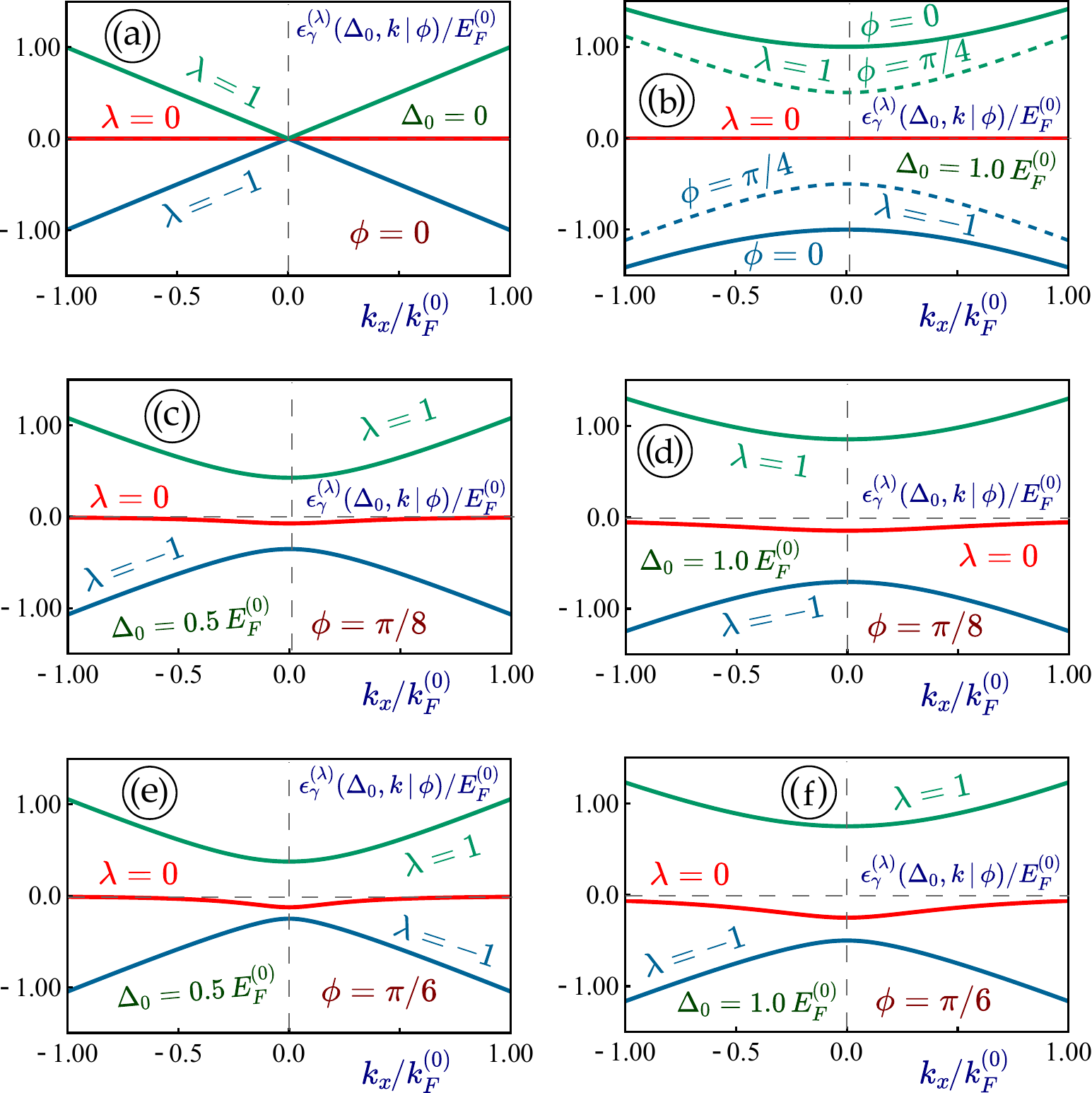}
\caption{(Color online) 
Energy dispersions for various $\alpha-\mc{T}_3$ materials in the presence of a finite energy gap $\Delta_0$. Panel $(a)$ demonstrates the bandstructure for graphene and an
arbitrary $\alpha-\mc{T}_3$ lattice for a zero gap $\Delta_0 = 0$, plot $(b)$ shows the energy subbands for graphene (solid lines) and a dice lattice $\phi = \pi/4$ 
(dashed lines) for a finite bandgap $\Delta_0 = 1.0\,E_F^{(0)}$. Panels $(c)$ and $(d)$ correspond to $\phi=\pi/8$, plots $(e)$ and $(f)$ - to  $\phi=\pi/6$.
Also, different panels demonstrate the energy bandstructure for different bandgaps $\Delta_0$, as labeled: $\Delta_0 = 0$ for panel $(a)$, $\Delta_0 = 0.5\,E_F^{(0)}$
- for plots $(c)$ and $(e)$, and  $\Delta_0 = 1.0\,E_F^{(0)}$ - for the left panels $(b)$ and $(d)$ and $(f)$.    
}
\label{FIG:1}
\end{figure}

Our main focus is, however, an $\alpha-\mc{T}_3$ material in the presence of a finite bandgap with the Hamiltonian obtained as

\begin{equation}
\label{hd}
\hat{\mc{H}}_\tau^{\,(\phi)}(\mbox{\boldmath$k$}) \longrightarrow \hat{\mc{H}}_\tau^{\,(\phi)}(\mbox{\boldmath$k$} \, \vert \, \Delta_0) = 
 \hat{\mc{H}}_\tau^{\,(\phi)}(\mbox{\boldmath$k$}) + \hat{\mc{H}}_a^{\,(\phi)}(\Delta_0) \, ,
\end{equation}
as it was employed in Ref.\,[\onlinecite{dey2019floquet}]. The additional $\phi$-dependent term $\hat{\mc{H}}^(\phi){\Delta_0}$ is  

\begin{equation}
\label{nh2}
\hat{\mc{H}}_a^{\,(\phi)}(\Delta_0)  = \frac{\Delta}{2} \, \hat{\Sigma}_z(\phi) \, 
= \Delta \, \left[
\begin{array}{ccc}
\cos^2 \phi & 0 & 0 \\
0 & - \cos2 \phi & 0 \\
0 & 0 & - \sin^2 \phi
\end{array}
\right] \ ,
\end{equation}
where $\hat{\hat{\Sigma}}_z^{\,(\phi)} = -i \, \left[\hat{\Sigma}_x(\phi), \hat{\Sigma}_y(\phi) \right]$ is derived and explained in Appendix 
\ref{apa}. Such a Hamiltonian with an additional $\phi$-dependent gap term \eqref{nh2} has been shown to describe an $\alpha-\mc{T}_3$ material irradiated  a circularly-polarized dressing field.

\medskip
\par 
We will now calculate the eigenenergies of Hamiltonian \eqref{hd}, but even before we do that it is clear that the two gap values at 
${\bf k} = 0$ depend on parameter $\alpha$ and that the middle band is no longer flat.

The energy dispersion for gapped $\alpha-\mc{T}_3$ described by Hamiltonian \eqref{hd} are obtained by 

\begin{eqnarray}
\label{ms1}
\varepsilon_\gamma(k,\Delta\,\vert\,\phi) &=& \frac{2}{\sqrt{3}} \, \sqrt{
k^2 + \frac{\Delta^2}{8} \left[
5 + 3 \cos (4 \phi)
\right] \,}\times \\
\nonumber
&& \times \cos \left\{ \frac{2 \pi (2 + \gamma)}{3} + \frac{1}{3} \,
\cos^{-1} \left( \frac{9 \sqrt{6}\, \Delta^3 \,\sin (2 \phi) \, \sin (4 \phi)}{\left\{
8 k^2 + \Delta^2[ 5 + 3 \cos (4 \phi) ]
\right\}^{3/2}
} \right) 
\right\}\ , 
\end{eqnarray}
where $\lambda = 0,\,1,\,2$ specifies three different energy bands. 

\par 
For a small bandgap $\Delta_0 \rightarrow 0$ and $k \neq 0$, subbands \eqref{ms1} are reduced to 

\begin{eqnarray}
\varepsilon_\gamma^{\,(\phi)} (k,\Delta \ll k) = \frac{2}{\sqrt{3}} \, \left\{
k + \frac{\Delta^2}{16 \, k} \left(
5 + 3 \cos4 \phi
\right)
\right\} \, \left[
\frac{\sqrt{3} \gamma}{2} + \frac{3 \sqrt{3} \Delta_0^3}{32 k^3} \right] \cdot \left(
2 - \vert \gamma \vert 
\right) 
\cdot \sin (2 \phi) \, \sin (4 \phi) \, . 
\end{eqnarray}

\medskip 

Once the energy eigenvalues have been calculated, the components of the wavefunction $\Psi^{(\gamma)}_{\lambda}$ $(\lambda = a, b, c)$, corresponding to each energy subband $\varepsilon^{\,(\phi)}_\gamma(k)$ could be found from the eigenvalue equation with Hamiltonian \eqref{hd} directly using

\begin{equation}
\label{tpsia}
\Psi^{\,\gamma}_{a}(k) = k \, \cos \phi \, \tet{e}^{- i \theta_{\bf k}} \, \left(  
\varepsilon_\lambda - \Delta_0 \, \cos^2 \phi
\right)^{-1} \, \Psi^{(\lambda)}_{b}(k)
\end{equation}

and 

\begin{equation}
\label{tpsic}
\Psi^{\,\gamma}_{a}(k) = k \, \sin \phi  \, \tet{e}^{i \theta_{\bf k}} \, \left(  
\varepsilon_\lambda + \Delta_0 \, \sin^2 \phi
\right)^{-1} \, \Psi^{(\lambda)}_{b}(k) \, . 
\end{equation}

The remaining component $\Psi^{(\lambda)}_{b}(k)$ could be now found from the normalization condition $\varepsilon_\gamma^{\,(\phi)} (k,\Delta \ll k)$

\begin{equation}
\label{tpsib}
\Psi^{\,\gamma}_{a}(k) = \left\{ 1 + 
\left[\frac{ k \, \cos \phi }{\left(  
\varepsilon_\lambda + \Delta_0 \, \cos^2 \phi
\right)} \right]^2 + 
\left[\frac{ k \, \sin \phi }{\left(  
\varepsilon_\lambda + \Delta_0 \, \sin^2 \phi
\right)} \right]^2
\right\}^{-1/2} \, . 
\end{equation}

Relations \eqref{tpsia} and \eqref{tpsic} are not applicable and need to be modified if both bandgap and the energy eigenvalue are zero. 
In this case, there is a direct relation between $\Psi^{\,\gamma}_{a}(k)$ and $\Psi^{\,\gamma}_{c}(k)$. Similar results for the energy dispersions and wavefunction of the gapped $\alpha-\mc{T}_3$ have been shown in Refs.~[\onlinecite{dey2019floquet}] and [\onlinecite{weekes2021generalized}]. Some details of the derivations of Eqs.~\eqref{tpsia},\eqref{tpsic} and \eqref{tpsib} are provided in our Appendix \ref{apa}. 

\par 
The energy dispersions  are presented in Fig.~\ref{FIG:1}. We clearly see that the middle band becomes dispersive, i.e., is no longer flat, once a finite bandgap is present. The middle band completely lies below zero-energy level and has its lowest point (negative peak) at ${\bf k} = 0$. The valence and conduction band dispersions $\varepsilon_\gamma^{\,(\phi)} (k,\Delta \ll k)$ receive non-equivalent bandgap which makes the whole bandstructure a lot more complex and less symmetric which is expected to lead to unique results for the optical conductivity. The only two cases 
where the middle band remains flat are graphene with $\phi = 0$ and a dice lattice of $\phi = \pi/4$ shown in panel $(b)$ of Fig.~\ref{FIG:1}.

\medskip
Both energy dispersions and the wavefunctions for the electronic states corresponding to different bands are important for calculating the optical conductivity. While the allowable energies mostly specify all the possible electron transitions between the occupied and free states and 
the regions for which the optical conductivity is nonzero, eigenstates components \eqref{tpsia},\eqref{tpsic} and \eqref{tpsib} are playing a substantial role in determining the exact frequency dependence of the optical conductivity.

\section{Optical conductivity} 
\label{sec3}

Once the electronic states and their dispersions are known, we are at position to calculate the dynamical optical conductivity of our lattice. According to the Kubo formalism, the real (absorptive) part of the longitudinal conductivity $\sigma_{T=0}^{\,(D)}(\omega \, \vert \, \Delta_0)$ is given by

\begin{equation}
\label{sxx}
\sigma_{T=0}^{\,(D)}(\omega \, \vert \, \Delta_0) = \frac{\pi e^2}{\hbar} \, \int \frac{d \xi}{\omega} \, \mbb{T}(\xi, \phi) \, \left[ f_0 (\xi + \omega \, \vert \, \mu(T), T ) - 
f_0 (\xi + \omega \, \vert \, \mu(T), T ) \right] \,  
\end{equation}
where where $g_s g_e = 4$ are the spin and valley degeneracy factors, $n_F (\xi\, \vert \, \mu, T)$ is the Fermi-Dirac distribution function and $\mu(T)$ 
is the chemical potential such that $\mu(T) \longrightarrow E_F$ and $n_F (\xi\, \vert \, \mu, T) \longrightarrow \Theta(E_F - \xi)$ for $T \rightarrow 0$. 

\medskip
The most crucial part of Eq.~\eqref{sxx} is the trace of the product of four matrices obtained as 

\begin{equation}
\label{t0}
\mbb{T}(\xi, \phi) =  \int \frac{d^2 k}{(2 \pi)^2}  \text{Tr} \left\{ 
\left[\hat{H}_0^{(\phi)},x \right]  \, \delta \left( \hat{H}_0^{(\phi)} - \xi - \omega \right) \,
\left[\hat{H}_0^{(\phi)},x \right]  \, \delta \left( \hat{H}_0^{(\phi)} - \xi \right) 
\right\} \, . 
\end{equation}

Operators $\left[\hat{H}_0^{(\phi)},x \right]$ could be rewritten using the definition of a current:

\begin{equation}
\label{jx}
\hat{j}_x = - i e \left[\hat{H}_\tau^{(\phi)}({\bf k}),x \right]= - e \frac{\pr \hat{H}_0^{(\phi)}({\bf k})}{\pr k_x} \, , 
\end{equation}
and a similar expression for current $\hat{j}_y$. Since we only consider isotropic materials and circularly polarized irradiation, we will only
calculate the $xx-$component of the optical conductivity.

\medskip 
\par 
Therefore, we also need to to evaluate 

\begin{equation}
\label{tr}
\mbb{T}(\xi, \phi) = \text{Tr} \left\{ 
\frac{\pr \hat{H}_0^{(\phi)}({\bf k})}{\pr k_x}  \, \delta \left( \hat{H}_0^{(\phi)} - \xi - \omega \right) \,
\frac{\pr \hat{H}_0^{(\phi)}({\bf k})}{\pr k_x}  \, \delta \left( \hat{H}_0^{(\phi)} - \xi \right) 
\right\} \, ,
\end{equation}
which is only possible in the representation in which Hamiltonian $\hat{H}_0^{(\phi)}({\bf k})$ is a diagonal matrix. In order to do that, we need to apply the following transformation

\begin{equation}
\frac{\pr \hat{H}_0^{(\phi)}({\bf k})}{\pr k_x}   \longrightarrow \left\langle \hat{P}_\psi^{(-1)}  \Big| \frac{\pr \hat{H}_0^{(\phi)}({\bf k})}{\pr k_x}  \Big|  \hat{P}_\psi \right\rangle
\end{equation}

and
\begin{eqnarray}
\nonumber
&& \delta \left( \hat{H}_0^{\,(\phi)}({\bf k}, \Delta_0)- \xi \right)  \longrightarrow  \left\langle \hat{P}_\psi^{(-1)}  \Big| \delta \left( \hat{H}_0^{(\phi)} - \varepsilon \right)  \Big|  \hat{P}_\psi \right\rangle = \left\{ 
\begin{array}{ccc}
\delta \left[ \epsilon^{\,(\phi)}_{\gamma = -1}(k) - \xi \right] & 0 & 0 \\
0 &  \delta \left[ \epsilon^{\,(\phi)}_{\gamma = 0}(k) - \xi \right]  & 0 \\
0 & 0 & \delta \left[ \epsilon^{\,(\phi)}_{\gamma = 1}(k) - \xi \right] 
\end{array}
\right\}
 \, ,
\end{eqnarray}
and

\begin{eqnarray}
&& \delta \left( \hat{H}_0^{(\phi)} - \varepsilon - \omega \right)   \longrightarrow  \left\{ 
\begin{array}{ccc}
\delta \left[ \epsilon^{\,(\phi)}_{\gamma = -1}(k, \Delta_0) - \xi- \omega \right] & 0 & 0 \\
0 &  \delta \left[ \epsilon^{\,(\phi)}_{\gamma = 0}(k, \Delta_0) - \xi - \omega \right]  & 0 \\
0 & 0 & \delta \left[ \epsilon^{\,(\phi)}_{\gamma = 1}(k, \Delta_0) - \xi - \omega \right] 
\end{array}
\right\}
 \, ,
\end{eqnarray}

The diagonalizing transformation matrix $\hat{P}_\psi$ is build from the component of the eigenfunction 

\begin{eqnarray}
\hat{P}^{\,(\phi)}_\psi({\bf k}) = \left\{ 
\begin{array}{ccc}
\Psi^{\,(\phi)}_{\gamma = -1} ({\bf k})_{[1]} & \Psi^{\,(\phi)}_{\gamma = 0} ({\bf k})_{[1]} & \Psi^{\,(\phi)}_{\gamma = 1} ({\bf k})_{[1]} \\
\Psi^{\,(\phi)}_{\gamma = -1} ({\bf k})_{[2]} & \Psi^{\,(\phi)}_{\gamma = 0} ({\bf k})_{[2]} & \Psi^{\,(\phi)}_{\gamma = 1} ({\bf k})_{[2]} \\
\Psi^{\,(\phi)}_{\gamma = -1} ({\bf k})_{[3]} & \Psi^{\,(\phi)}_{\gamma = 0} ({\bf k})_{[3]} & \Psi^{\,(\phi)}_{\gamma = 1} ({\bf k})_{[3]}
\end{array}
\right\}
\end{eqnarray}

This transformation obviously does not change a trace of the product of four matrices $\hat{A}^{\,(\phi)}({\bf k}) \times \hat{\mbb{O}}^{\,(\phi)}_\epsilon({\bf k}) \times \hat{A}^{\,(\phi)}({\bf k}) \times \hat{\mbb{O}}^{\,(\phi)}_{\xi + \omega}({\bf k})$ since 

\begin{eqnarray}
&& \text{Tr} \left\{  
\left[  \hat{P}_\psi^{(-1)} \times  \hat{A} \times \hat{P}_\psi \right] 
\times 
\left[  \hat{P}_\psi^{(-1)} \times  \hat{\mbb{O}}_{\xi} \times \hat{P}_\psi \right] 
\times \left[  \hat{P}_\psi^{(-1)} \times   \hat{A} \times \hat{P}_\psi \right]  
\times 
\left[  \hat{P}_\psi^{(-1)} \times  \hat{\mbb{O}}_{\xi + \omega} \times \hat{P}_\psi \right] \right\}  =  \\
\nonumber
&&   = \text{Tr} \left\{  \hat{P}_\psi^{(-1)} \times  \hat{A} \times \hat{\mbb{O}}_{\xi} \times \hat{A} \times \hat{\mbb{O}}_{\xi + \omega} \times \hat{P}_\psi \right\}  = \\
\nonumber 
&&  = \text{Tr} \left\{ \hat{A} \times  \hat{\mbb{O}}_{\xi} \times \hat{A} \times \hat{\mbb{O}}_{\xi+ \omega} \times \hat{P}_\psi \times \hat{P}_\psi^{(-1)} 
\right\}  =  \text{Tr} \left\{ \hat{A} \times  \hat{\mbb{O}}_{\xi} \times \hat{A} \times \hat{\mbb{O}}_{\xi + \omega}  \right\} \, .
\end{eqnarray} 
where $\hat{A}^{\,(\phi)}({\bf k}) = \pr \hat{H}_0^{(\phi)}({\bf k})/ (\pr k_x)$, $\hat{\mbb{O}}_{\epsilon}({\bf k}) = \delta \left[ \hat{H}_0^{(\phi)} - \varepsilon \right]$ and $\hat{\mbb{O}}_{\xi + \omega} ({\bf k}) = \delta \left[ \hat{H}^{\,(\phi)}_0({\bf k}) - \xi - \omega \right]$. Here, we also used a relation $\text{Tr} \left\{\hat{A} \times \hat{B} \right\} = \text{Tr} \left\{ \hat{B} \times \hat{A} \right\}$.

\medskip 

For a gapless $\alpha-\mc{T}_3$, the transformation matrix is simplified to 

\begin{eqnarray}
\hat{P}_\psi^{\,(\phi)}({\bf k}) = \frac{1}{\sqrt{2}} \, \left\{ 
\begin{array}{ccc}
\tet{e}^{-i  \theta_{\bf k}} \, \cos \phi & \sqrt{2} \, \tet{e}^{-i  \theta_{\bf k}} \, \sin \phi  & \tet{e}^{-i  \theta_{\bf k}} \, \cos \phi  \\
1 & 0 & -1 \\
\tet{e}^{i  \theta_{\bf k}} \, \sin \phi & \sqrt{2} \, \tet{e}^{i  \theta_{\bf k}} \, \cos \phi & \tet{e}^{i  \theta_{\bf k}} \, \sin \phi
\end{array}
\right\}
\end{eqnarray}

and for graphene

\begin{eqnarray}
\hat{P}_\psi^{(0)}({\bf k}) = \frac{1}{\sqrt{2}} \, \left\{ 
\begin{array}{cc}
1 & 1 \\
\tet{e}^{i  \theta_{\bf k}} & - \tet{e}^{i  \theta_{\bf k}}
\end{array}
\right\}
\end{eqnarray}
which is a unitary matrix. This technique in principle leads to an analytical result for the trace \eqref{t0}, as well as the corresponding optical conductivity. However, for a gapped $\alpha-\mc{T}_3$ material the obtained expressions are too lengthy and complicated so that this calculation approach is very convenient and is often 
use for a numerical evaluation of $\sigma_{T=0}^{\,(D)}(\omega \, \vert \, \Delta_0)$.

\begin{figure} 
\centering
\includegraphics[width=0.99\textwidth]{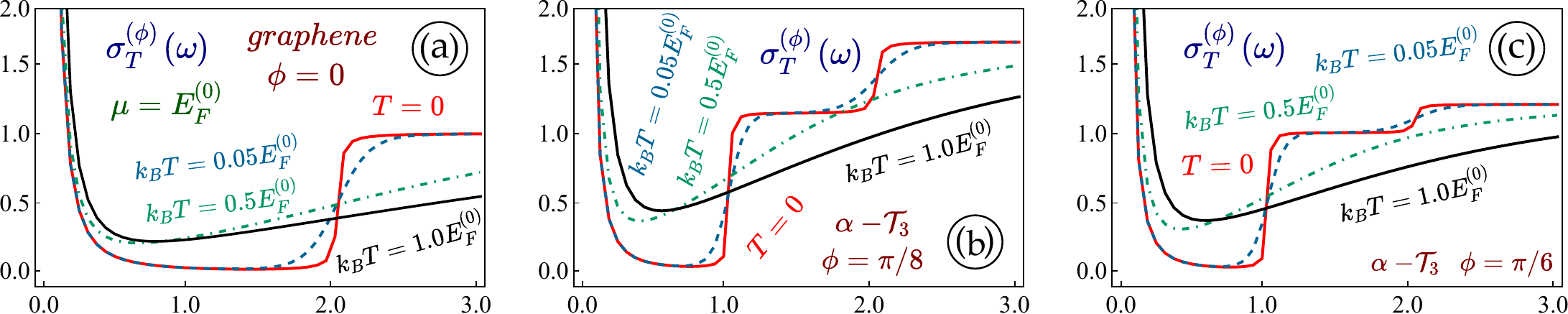}
\caption{(Color online) Temperature-dependent longitudinal optical conductivity for graphene and different $\alpha-\mc{T}_3$ materials with a finite chemical potential
$\mu = 1.0\,E_F^{(0)}$. Panel $(a)$ shows the photon frequency dependence of the optical conductivity for graphene with $\phi = 0$, plots $(b)$ and $(c)$ - for $\alpha-\mc{T}_3$ materials with $\phi = \pi/8$ and $\phi = \pi/6$, correspondingly. Red and solid curve is used to describe the zero-temperature result, blue and dashed is related to $k_B T = 0.05\,E_F^{(0)}$, green and dash-dotted line - to $k_B T = 0.5\,E_F^{(0)}$ and the black solid curve corresponds to $k_B T = 1.0\,E_F^{(0)}$ on each panel.}
\label{FIG:2}
\end{figure}

\medskip
\par
\begin{figure} 
\centering
\includegraphics[width=0.6\textwidth]{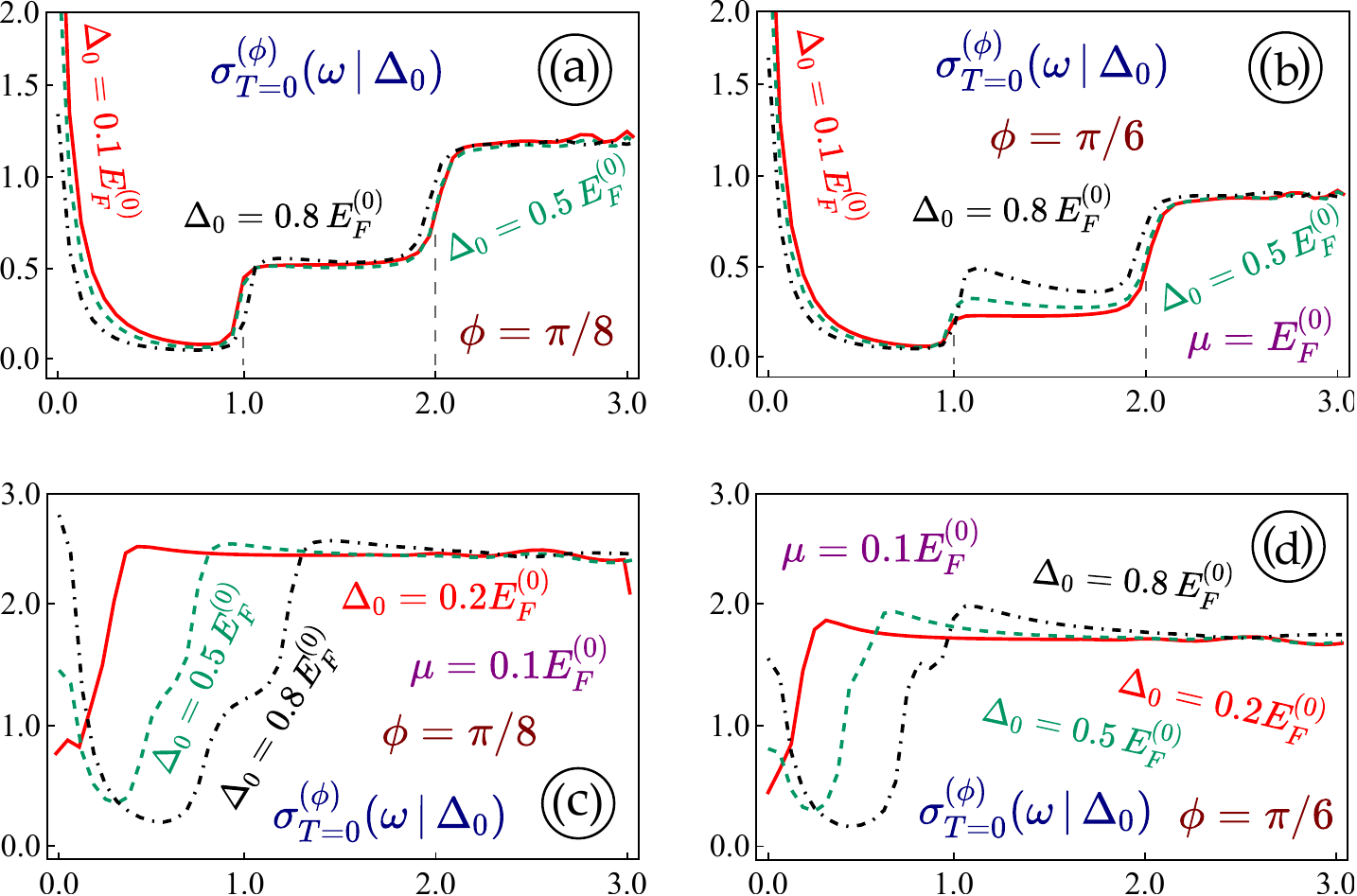}
\caption{(Color online) Longitudinal optical conductivity for various $\alpha-\mc{T}_3$ materials in the presence of an energy bandgap $\Delta_0$. Top panels 
$(a)$ and $(b)$ correspond to a finite chemical potential $\mu = 1.0\,E_F^{(0)}$, while the lower plots $(c)$ and $(d)$ - to its nearly-zero level 
$\mu = 0.1\,E_F^{(0)}$ just above the middle band. The two left panels $(a)$ and $(c)$ show the results for $\phi  = \pi/8$, while the right ones $(b)$ and $(d)$ are related to $\phi  = \pi/6$. All the panels are shown for zero temperature.}
\label{FIG:3}
\end{figure}

\medskip
\par
\begin{figure} 
\centering
\includegraphics[width=0.6\textwidth]{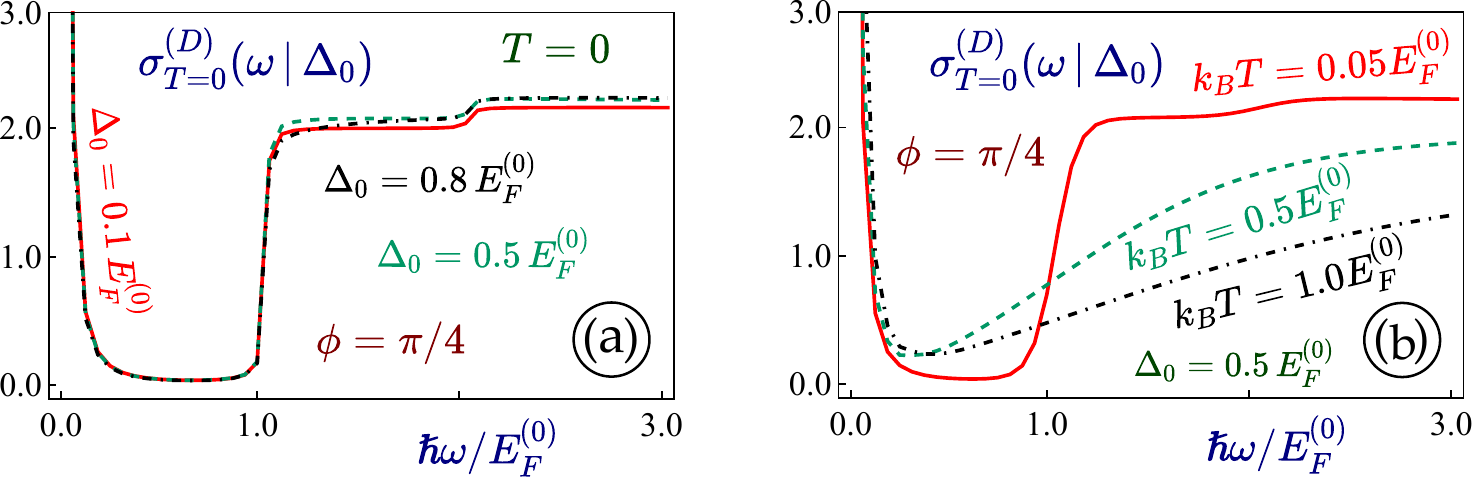}
\caption{(Color online) Temperature-dependent longitudinal optical conductivity for a dice lattice with $\phi = \pi/4$ in the presence of an energy bandgap $\Delta_0$. Panel $(a)$ demonstrates the zero-temperature conductivity $\sigma_{T=0}^{\,(D)}(\omega \, \vert \, \Delta_0)$ for different values of the bandgap $\Delta_0$, as labeled. Plot $(b)$ represents the conductivity $\sigma_{T}^{\,(D)}(\omega \, \vert \, \Delta_0)$ for a fixed gap 
$\Delta_0 = 0.5\,E_F^{(0)}$ at various temperatures: red and solid curve is used to describe the case with $k_B T = 0.05\,E_F^{(0)}$, green and dashed is related to $k_B T = 0.5\,E_F^{(0)}$, black and dash-dotted line - to $k_B T = 1.0\,E_F^{(0)}$.}
\label{FIG:4}
\end{figure}

Another well-known technique of calculating the dynamical optical conductivity of a two-dimensional lattice is based on finding the spectral function
which is connected to the Green's function of an electron in a gapped $\alpha-\mc{T}_3$ lattice as 

\begin{equation}
\label{def1}
\hat{\mc{G}}^{\,(\phi)} (k,\xi \, \vert \, \Delta_0) = \int \frac{d \varepsilon}{2 \pi (\xi - \epsilon)} \, \hat{\mbb{S}}^{\,(\phi)} (k, \epsilon \, \vert \, \Delta_0) \, . 
\end{equation}
Applying this approach, we would not have to transfer the velocity operators to the representation which diagonalizes the Hamiltonian. While the two approached 
clearly lead to the same results, calculation of the Green's function and spectral function of an $\alpha-\mc{T}_3$ material and a dice lattice represent some independent interest.   

\medskip 
The spectral functions for a gapped dice lattice and an arbitrary $\alpha-\mc{T}_3$ materials are obtained in Appendices \ref{apb} and \ref{apc}, correspondingly. Once the elements of the greens function are decomposed into partial fraction, equation \eqref{def1} will be used directly as $1/(\xi - \epsilon_0) \longrightarrow
2 \pi \, \delta (\xi - \epsilon_0)$. However, such a partial fraction decomposition is not always straightforward if a denominator represents a cubical equation.
 
%
%

%

\par
The trace term in Eq.~\eqref{t0} is now obtained as 

\begin{equation}
\label{trace2B}
\mbb{T}^{\,(\phi)}(\xi, \omega \, \vert \, \Delta_0) = \int \frac{d^2 {\bf k}}{2 \pi} \, \text{Tr} \left\{ 
\hat{V}_x^{\,(\phi)} \times \hat{\mbb{S}}^{\,(\phi)} (k, \xi + \omega\, \vert \, \Delta_0) \times \hat{V}_x^{\,(\phi)} \times 
\hat{\mbb{S}}^{\,(\phi)} (k, \xi \, \vert \, \Delta_0)
\right\} \, .
\end{equation}

The Green's function is explicitly obtained as 

\begin{equation}
\hat{\mc{G}}^{\,(\phi)} (k,\xi \, \vert \, \Delta_0)  = \left\{ 
\begin{array}{ccc}
\mc{G}^{\,(\phi)}_{1\,[1, 0]} (k, \xi)  & - \cos \phi \, \tet{e}^{ - i \theta_{\bf k}} \, \mc{G}_{2} (k, \xi) 
 & - \sin (2 \phi) \, \tet{e}^{ - 2 i \theta_{\bf k}} \, (k/2 z) \, \mc{G}_{2} (k, \xi) \\
- \cos \phi \, \tet{e}^{i \theta_{\bf k}} \mc{G}_{2} (k, \xi) & \mc{G}_{1\,[0, 1]} (k, \xi)  & 
- \sin \phi \, \tet{e}^{ - i \theta_{\bf k}} \, \mc{G}_{2} (k, \xi)  \\
- \sin (2 \phi) \, \tet{e}^{2 i \theta_{\bf k}} \, (k/2 z) \, \mc{G}_{2} (k, \xi) & 
- \sin \phi \, \tet{e}^{i \theta_{\bf k}} \, \mc{G}_{2} (k, \xi) & \mc{G}^{\,(\phi)}_{1\,[-1, 1]} (k, \xi) 
\end{array}
\right\} \, ,
\end{equation}
where

\begin{eqnarray}
&& \mc{G}^{\,(\phi)}_{1\,[\beta, \nu]} (k, \xi) = \frac{z^2 + \beta \, k^2 \, (\nu - \sin^2 \phi)}{z^3 - k^2 z} \, , \\ 
\nonumber
&& \mc{G}_{2} (k, \xi) = \frac{k}{z^2 - k^2} \, .
\end{eqnarray}
The current operator defined in \eqref{jx} and $\hat{{\bf j}}_{\tau,s}^{\,(\phi)} = - e \hat{{\bf V}}_{\tau,s}^{\,(\phi)}$. Therefore, we obtain 

%

\begin{equation}
\label{vx}
\hat{V}^{\,(\phi)}_x = v_F \, \hat{\Sigma}^{\,(\phi)}_x  = \left\{ 
\begin{array}{ccc}
0 & \cos \phi & 0 \\
\cos \phi & 0 &  \sin \phi\\
0 & \sin \phi & 0
\end{array}
\right\}
\end{equation}

and 

\begin{equation}
\label{vy}
\hat{V}_y = v_F \, \hat{\Sigma}^{\,(\phi)}_y = \left\{ 
\begin{array}{ccc}
0 & - i \cos \phi & 0 \\
i \cos \phi & 0 &  - i \sin \phi\\
0 & i \sin \phi & 0
\end{array}
\right\}
\end{equation}

which does not depend on the energy bandgap, i.e., Eqs.~\eqref{vx} and \eqref{vy} are the same for the cases of zero and a finite energy gap in $\alpha-\mc{T}_3$.

\medskip
Thus, the trace in Eq.~\eqref{trace2B} for a gapped dice lattice is obtained in Appendix \ref{apb} as

\begin{equation}
\label{MtraceA}
\mbb{T}(k, \xi \, \vert \, \omega)= \frac{1}{2} \, \int\limits \frac{d^2 {\bf k}}{2 \pi} \, \left[ \mbb{T}_1(k, \xi \, \vert \, \omega) + 
\, \mbb{T}_2 (k,\xi \, \vert \, \omega)
\right] \, , 
\end{equation}

where

\begin{eqnarray}
&& \mbb{T}_1(k , \xi \, \vert \, \omega) = \left[\mbb{S}^\phi_{21} (k,\xi) + \mbb{S}^\phi_{23} (k,\xi) \right] \times
\left[\mbb{S}^\phi_{21} (k,\xi + \omega) + \mbb{S}^\phi_{23} (k,\xi + \omega) \right] + \\
\nonumber
&& + \left[\mbb{S}^\phi_{12} (k,\xi) + \mbb{S}^\phi_{32} (k,\xi) \right] \times
\left[\mbb{S}^\phi_{12} (k,\xi + \omega) + \mbb{S}^\phi_{32} (k,\xi + \omega) \right]
\\
\nonumber 
&& \mbb{T}_2(k , \xi \, \vert \, \omega) =\mbb{S}^\phi_{22} (k,\xi + \omega)  \times \left[\mbb{S}^\phi_{11} (k,\xi) + \mbb{S}^\phi_{33} (k,\xi) \right] + 
\mbb{S}^\phi_{22} (k,\xi)  \times \left[\mbb{S}^\phi_{11} (k,\xi + \omega) + \mbb{S}^\phi_{33} (k,\xi + \omega) \right] 
\end{eqnarray}

Specifically, we look at the zero-gap limit of a dice lattice

\begin{equation}
\label{MtraceA2}
\mbb{T}^{\,(D)}_{\Delta_0 = 0}(k, \xi \, \vert \, \omega)= \int\limits \frac{d^2 {\bf k}}{4 \pi} \, \left[ \mbb{T}^{\,(D)}_{1,\Delta_0 = 0}(k, \xi \, \vert \, \omega) + \mbb{T}^{\,(D)}_{2,\Delta_0 = 0}(k, \xi \, \vert \, \omega) + 4 \mbb{T}^{\,(D)}_{3,\Delta_0 = 0}(k, \xi \, \vert \, \omega)
\right] \, , 
\end{equation}
where

\begin{eqnarray}
&& \mbb{T}^{\,(D)}_{1,\Delta_0 = 0}(k, \xi \, \vert \, \omega) = \delta(\omega + \xi - k) \times \left[
5 \, \delta(\xi - k ) - 3 \, \delta(\xi +k) \right]
\, , \\ 
\nonumber 
&& \mbb{T}^{\,(D)}_{2,\Delta_0 = 0}(k, \xi \, \vert \, \omega) = \delta(\omega + \xi + k) \times \left[
5 \, \delta(\xi + k ) - 3 \, \delta(\xi - k) \right]
\, , \\ 
\nonumber
&& \mbb{T}^{\,(D)}_{3,\Delta_0 = 0}(k, \xi \, \vert \, \omega) = \delta(\xi) \times \left[
 \delta(\xi +\omega - k) + \delta(\xi +\omega + k)
\right] \, . 
\end{eqnarray}

\begin{equation}
\label{Mt}
\mbb{T}^{\,(D)}_{\Delta_0 = 0}(k, \xi \, \vert \, \omega)= \int\limits \frac{d^2 {\bf k}}{4 \pi} \, \left[ \mbb{T}^{\,(D)}_{1,\Delta_0 = 0}(k, \xi \, \vert \, \omega) + 4 \mbb{T}^{\,(D)}_{3,\Delta_0 = 0}(k, \xi \, \vert \, \omega)
\right] \, , 
\end{equation}
where

\begin{eqnarray}
&& \mbb{T}^{\,(D)}_{1,\Delta_0 = 0}(k, \xi \, \vert \, \omega) = \sum \limits_{\beta = \pm 1} \delta(\omega + \xi - \beta k) \, 
\times \left[
5 \, \delta(\xi -   \beta k ) - 3 \, \delta(\xi +  \beta k) \right]
\, , \\ 
\nonumber 
&& \mbb{T}^{\,(D)}_{1,\Delta_0 = 0}(k, \xi \, \vert \, \omega) = \delta(\xi) \times \sum \limits_{\beta = \pm 1}
 \delta(\xi +\omega + \beta k) 
 \, . 
\end{eqnarray}

The $\Delta_0 \longrightarrow 0$ limit of a dice lattice is also a $\phi \longrightarrow \pi/4$ limit of a general $\alpha-\mc{T}_3$ material whose optical conductivity

For our final calculation of the optical conductivity, we use a Lorentzian representation of the delta function

\begin{equation}
\label{Mdeltarep}
\delta_\rho (x) = \frac{1}{\pi} \, \frac{\rho}{\rho^2 + x^2}
\end{equation}
with a small but yet a finite broadening parameter $\rho = 0.01\,E_F^{(0)}$ in order to account for the impurity scattering $\backsim 2 \rho$ which enables more  realistic results for the optical conductivity.

\medskip 
The behavior of optical conductivity of graphene and $\alpha-\mc{T}_3$ without a gap, shown in Fig:~\ref{FIG:2}, are have been known to reveal two jumps of $\sigma_{T=0}^{\,(\phi)}(\omega)$ at $\mu$ and $2\mu$ at the level of $\cos(2 \phi)^2$ and $2 \sin(2 \phi)^2$, correspondingly. All the optical transitions below the Fermi level are blocked while the transitions in the range $\mu < \omega < 2 \mu$ are allowable only because of the presence of the flat band. The limitations on the electron transitions are lifted it the temperature becomes finite. By $\mu$ or $\mu(T)$ we mean a general chemical potential which is equal to the Fermi energy at $T=0$. Our Fermi energy is taken either $0.1\,E_F^{(0)}$ or $1,0\,E_F^{(0)}$, where our unit of energy $E_F^{(0)} =  50 \,meV$ is a typical Fermi energy in graphene, corresponding to electron density $n_e = 10^{11}\,cm^{-2}$.

\par 
The presence of another energy band enables such transitions from the flat band to the conduction band and vice versa. Its affect on the electron states is solely determined by phase $\phi$ (or parameter $\alpha$). Therefore, the ratio between the two abrupt jumps of the optical conductivity is decreasing for increasing $\phi$. Our main interest for gapless $\alpha-\mc{T}_3$ in the temperature dependence which leads to a gradual reduction of the sharp jumps and a finite $\sigma_{T}^{\,(\phi)}(\omega)$ for all frequencies and comparing this behavior with the case of a finite flat band.

\medskip
\par 
Our key results are related to the optical conductivity of an $\alpha-\mc{T}_3$ in the presence of a finite bandgap, shown in Fig.~\ref{FIG:3}. We see that the 
obtained frequency dependence of the optical conductivity combines the main feature of that in regular $\alpha-\mc{T}_3$, including two-step abrupt increase of 
$\sigma_T^{\,(\phi)}(\omega)$ for $T=0$, and the optical conductivity in silicene with a finite spin-orbit gap in which we see a sharp peak and a later drop of $\sigma_T^{\,(\phi)}(\omega)$ with increasing $\omega$ as $\backsim \left(1 - \Delta_0^2/\omega^2  \right)$.

\par 
A very special situation occurs when the Fermi level $\mu$ is located below the conduction band in the gap region.  The frequency dependence of the optical  conductivity in this case does not show intraband Drude conductivity peak $\backsim \delta(\omega)/\omega$ for $\omega \longrightarrow 0$, which we observed for all other cases. For an absent or a small bandgap, $\sigma_T^{\,(\phi)}(\omega)$ remains nearly flat for almost all frequencies, similarly to what was previously found in silicene.\,\cite{tabert2015electronic} Thus, the lowest frequency with a finite optical absorption is given by either $\mu$ or $2 \delta$, whichever is larger.

\medskip
\par
Finally, we put a main focus on the case of a gapped dice lattice in which the flat band remains flat even if the valence and conduction bands receive a gap. Interestingly, here we only see a one large jump of the optical conductivity at $\mu = 1.0 \,E_F^{(0)}$ and apart from a small extra step, we see almost no dependence of $\sigma_T^{\,(\phi)}(\omega)$ on $\omega$ above the Fermi level. We have also found that this only jump of the optical conductivity has no further decreasing dependence for $\omega > \mu$. For gapped $\alpha-\mc{T}_3$ shown in Fig.~\ref{FIG:3} $(b)$, only the first peak shows a decrease in this frequency range. It is also completely missing for a gapped dice lattice, presented in Fig.~\ref{FIG:4} for zero and finite temperatures.

\section{Summary and Remarks} 
\label{sec4}

\medskip 

We have calculated the dynamical optical conductivity for various kinds of $\alpha-\mc{T}_3$ materials in the presence of a finite energy bandgap for different values of electron doping, including the final temperatures.  The bandstructure of a gapped $\alpha-\mc{T}_3$ lattice demonstrates some very unusual properties since except for the two limiting cases of graphene $\phi = 0$ and a dice lattice  $\phi = \pi/4$, the middle (initially, flat) band receives a finite dispersion. Therefore, it could be partially or completely doped. Also, it is important that the infinite degeneracy of the flat band is lifted and the Fermi level could be placed within its energy range. Such energy dispersions and the electronic states could be obtained as a result of the interaction between Dirac electrons in an $\alpha-\mc{T}_3$ material with off-resonance dressing field.

\medskip 

We have found that the dynamical, or frequency-dependent, optical conductivity shares all the known features of that in a zero-gap $\alpha-\mc{T}_3$ materials and silicene with two inequivalent bandgaps. At zero temperature, the optical conductivity in graphene for $\omega < 2 \mu$ is equal to zero because of the Pauli blocking, meaning that the carriers cannot transfer between the two occupied states. 

\par 
For $\alpha-\mc{T}_3$, the  conductivity shows two separate steps corresponding to the transitions between the valence and conduction bands, and the flat and conduction bands, correspondingly. For a finite Fermi level, we also observe an infinite Drude conductivity peak $\backsim \delta(\omega)/\omega$ at $\omega \longrightarrow 0$ due to the  intraband electron transitions with vanishing energy transfer between the initial and final electron states.

\medskip

In a specific case of gapped $\alpha-\mc{T}_3$ materials, only one of the two peaks of the optical conductivity shows a $\backsim \left(1 - \Delta_0^2/\omega^2  \right)$ decrease which is not observed for a dice lattice with only one such jump of optical conductivity for $\omega = \mu$. The latter case is special because the flat band remains unaffected by the external radiation. Thus, our results are simplified and an analytical solution for the Green’s function, spectral function and the optical conductivity have been obtained. We have also derived analytical expressions for the zero-bandgap limit of a dice lattice which also corresponds to $\phi = \pi/4$ limit of a gapless $\alpha-\mc{T}_3$ material.

\medskip 

Optical conductivity is one of the most important measurable quantities which has been thoroughly studied for any new low-dimensional material. Its unique properties and, specifically, its frequency dependence allows for an experimental verification of the bandstructure of a material for various doping levels. \,\cite{grazianetti2018optical} We are confident that our unusual results for the optical conductivity of a gapped $\alpha-\mc{T}_3$ could be undoubtedly employed in a number of promising technical applications.

\appendix

\section{Appendix A: Energy dispersions and wavefunctions for gapped $\alpha-\mc{T}_3$}
\label{apa}

Hamiltonian \eqref{mainH} could be constructed using the following two $\phi$-dependent $3 \times 3$ matrices $\hat{\mbox{\boldmath$S$}}(\phi)= \left\{  \hat{S}_x(\phi),\,\hat{S}_y(\phi) \right\}$, where

\begin{equation}
\label{Sxp}
\hat{\Sigma}_x(\phi) = \left[
\begin{array}{ccc}
 0 & \cos \phi & 0 \\
 \cos \phi & 0 & \sin \phi \\
 0 & \sin \phi & 0
\end{array}
\right] \ ,
\end{equation}

and

\begin{equation}
\label{Syp}
\hat{\Sigma}_y(\phi) = i \,\left[
\begin{array}{ccc}
 0 & -\cos \phi & 0 \\
 \cos \phi & 0 & -\sin \phi \\
 0 & \sin \phi & 0
\end{array}
\right] \ .
\end{equation}

Matrices \eqref{Sxp} and \eqref{Syp} are a $\phi$-dependent generalization of $3 \times 3$ Pauli matrices 

\begin{equation}
\hat{\Sigma}^{(3)}_{x} = \frac{1}{\sqrt{2}} \, \left[
\begin{array}{ccc}
 0 & 1 & 0 \\
 1 & 0 & 1 \\
 0 & 1 & 0
\end{array}
\right] \ ,
\label{sig1}
\end{equation}

\begin{equation}
\hat{\Sigma}^{(3)}_{y} = \frac{i}{\sqrt{2}} \, \left[
\begin{array}{ccc}
 0 & -1 & 0 \\
 1 & 0 & -1 \\
 0 & 1 & 0
\end{array}
\right] \ ,
\label{sig2}
\end{equation}
which are related to a dice lattice with $\phi = \pi/4$ is taken. On the other hand, for $\phi \rightarrow 0$, matrices in Eqs.\,\eqref{Sxp} and \eqref{Syp} reduce to regular spin-$1/2$ Pauli matrices used to describe the Hamiltonian in graphene.

\par
We also need to employ another Pauli matrix $\hat{\Sigma}^{(3)}_{z}$, defined as  

\begin{equation}
\label{szdice}
\hat{\Sigma}^{(3)}_{z} = \left[
\begin{array}{ccc}
 1 & 0 & 0 \\
 0 & 0 & 0 \\
 0 & 0 & -1
\end{array}
\right] \ , 
\end{equation}
indented to represent a gap term in pseudospin-$1$ Hamiltonian \,\eqref{mainH}.

\medskip
\par

We will now calculate the eigenenergies of Hamiltonian \eqref{hd}, but even before we do that it is clear that the two gap values at 
${\bf k} = 0$ depend on parameter $\alpha$ and that the middle band is no longer flat. 

\medskip

The energy sought energy dispersions are obtained from the following secular equation

\begin{equation}
\label{ed2}
\varepsilon^{\,(\phi)}(k)^3 - c_1 \, \varepsilon^{\,(\phi)}(k)  - c_0 = 0 \ ,
\end{equation}

where 

\begin{eqnarray}
&& c_1 = k^2 + \frac{\Delta_0^2}{8} \left[
5 + 3 \cos (4 \phi) 
\right] \, ,\\
\nonumber
&& c_0 = \left( \frac{\Delta_0}{2} \right)^3 \, \sin(2 \phi) \,  \sin (4 \phi)  \, . 
\end{eqnarray}
We immediately see the flat band remains dispersionless if $c_1 = 0$ which is achievable for $\phi = 0$ and $\phi = \pi/4$ - the two marginal cases of 
graphene and a dice lattice. For all other $0 < \alpha < 1$ the flat band will be curved for a finite bandgap.   

\medskip 
Mathematically, Eq.~\eqref{ed2} is a {\it depressed cubical equation} with a missing $\backsim \varepsilon^2$ term. There is a number of ways to 
solve such equations, but if we are aware that there are three real solutions to this equation, it seems the best to use Viete's solutions 
based on trigonometric functions.

\par
We want to rewrite Eq.~\eqref{ed2} to make it look similar to the following trigonometric identity

\begin{equation}
\cos^3 \theta - \frac{3}{4}\, \cos \theta - \frac{3}{4}\, \cos (3 \theta) = 0 \, ,
\end{equation}
or, alternatively, 

\begin{equation}
\cos \theta \, \left[ 
\cos^2 \theta - \cos^2 \left(\frac{\pi}{6} \right)
\right] = \frac{1}{4} \, \cos (3 \theta) \, , 
\end{equation}
which is achieved by a substitution 

\begin{equation}
\varepsilon^{\,(\phi)}(k) = c_1^{1/2} \, \frac{\cos \theta}{\cos (\pi/6)} \, . 
\end{equation}
As a result, we obtain

\begin{equation}
\cos (3 \theta) = 4 \cos^3 \left(\frac{\pi}{6} \right) \, \frac{c_0}{c_1^{3/2}} \, .  
\end{equation} 

Thus, the sought solutions for the energy eigenvalues are  

\begin{equation}
\varepsilon^{\,(\phi)}_\gamma (k) = \frac{c_1^{1/2}}{\cos (\pi/6)} \, \cos \left\{
\frac{2 \pi \gamma}{3} + \frac{1}{3} \, \cos^{-1} \left[ 
4 \cos^3 \left(\frac{\pi}{6} \right) \, \frac{c_0}{c_1^{3/2}}
\right] 
\right\} 
\end{equation}

gives rise to three solutions, given by

\begin{equation}
\label{as1}
\varepsilon^{\,(\phi)}_\gamma(k,\Delta\,\vert\,\phi)= \frac{2}{\sqrt{3}} \, \sqrt{
k^2 + \frac{\Delta^2}{8} \left(
5 + 3 \cos4 \phi
\right) \,}\,
\cos \left[ \frac{2 \pi (2+\gamma)}{3} + \frac{1}{3} \,
\cos^{-1} \left( \frac{9 \sqrt{6}\, \Delta^3 \,\sin 2 \phi \, \sin 4 \phi}{\left[
8 k^2 + \Delta^2(5 + 3 \cos (4 \phi))
\right]^{3/2}
} \right) 
\right] \ , 
\end{equation}
where $\gamma = 0,\,1,\,2$ specifies three different energy bands.

\medskip 
For a small bandgap $\Delta_0 \rightarrow 0$, subbands \eqref{as1} ($k \neq 0$) are simplified as 

\begin{eqnarray}
\varepsilon_\gamma(k,\Delta \ll k \,\vert\,\phi) = \frac{2}{\sqrt{3}} \, \left[
k + \frac{\Delta^2}{16 \, k} \left(
5 + 3 \cos4 \phi
\right)
\right] \times \left\{
\begin{array}{l}
\sqrt{3} \gamma/ 2 + 3 \sqrt{3} \Delta_0^3/(32 k^3) \cdot \sin (2 \phi) \, \sin (4 \phi) \,\,\,\text{for}\,\,\,\lambda = \pm 1\, , 
\\
3 \sqrt{3} \Delta_0^3/(16 k^3) \cdot \sin (2 \phi) \, \sin (4 \phi) \,\,\,\,\text{for}\,\,\,\,\lambda = 0\, .
\end{array}
\right.
\end{eqnarray}

\medskip

Once the energy eigenvalues have been calculated, the components of the wavefunction $\Psi^{\gamma}_{i}$, $(i = a, b, c)$, corresponding to each energy subband
$\varepsilon_\gamma(k)$ could be found from the main eigenvalue equation ... in a relatively simple way

\begin{equation}
\label{psib1}
\Psi^{(\gamma)}_{b}(k) = \frac{\tet{e}^{i \theta_{\bf k}}}{k \, \cos \phi} \, \left(  
\varepsilon_\gamma - \Delta_0 \, \cos^2 \phi
\right) \, \Psi^{(\gamma)}_{a}(k)
\end{equation}

or alternatively,

\begin{equation}
\label{psia1}
\Psi^{(\gamma)}_{a}(k) = k \, \cos \phi \, \tet{e}^{- i \theta_{\bf k}} \, \left(  
\varepsilon_\gamma - \Delta_0 \, \cos^2 \phi
\right)^{-1} \, \Psi^{(\gamma)}_{b}(k)
\end{equation}

and 

\begin{equation}
\label{psib2}
\Psi^{(\gamma)}_{b}(k) = \frac{\tet{e}^{- i \theta_{\bf k}}}{k \, \sin \phi} \, \left(  
\varepsilon^{\,(\phi)}_\gamma(k) + \Delta_0 \, \sin^2 \phi
\right) \, \Psi^{(\gamma)}_{c}(k)
\end{equation}

which is equivalent to

\begin{equation}
\label{psic2}
\Psi^{(\gamma)}_{c}(k) = k \, \sin \phi  \, \tet{e}^{i \theta_{\bf k}} \, \left(  
\varepsilon^{\,(\phi)}_\gamma(k) + \Delta_0 \, \sin^2 \phi
\right)^{-1} \, \Psi^{(\gamma)}_{b}(k)
\end{equation}

The remaining component $\Psi^{(\gamma)}_{b}(k)$ could be now found from the normalization condition

\begin{equation}
\Psi^{(\gamma)}_{b}(k) = \left\{ 1 + 
\left[\frac{ k \, \cos \phi }{\left(  
\varepsilon^{\,(\phi)}_\gamma(k) + \Delta_0 \, \cos^2 \phi
\right)} \right]^2 + 
\left[\frac{ k \, \sin \phi }{\left(  
\varepsilon^{\,(\phi)}_\gamma(k) + \Delta_0 \, \sin^2 \phi
\right)} \right]^2
\right\}^{-1/2}
\end{equation}

Relations \eqref{psib1} and \eqref{psib2} are not applicable and need to be modified if both bandgap and the energy eigenvalue are zero.

\section{Green's functions and spectral functions for a gapped dice lattice.}
\label{apb}

The elements $\mbb{S}_{ij}^{\,(D)} (k,\xi \, \vert \, \Delta_0)$ spectral function

\begin{equation} \hat{\mbb{S}}^{\,(D)} (k,\xi \, \vert \, \Delta_0) = \left\{
\begin{array}{ccc}
\mbb{S}_{11}^{\,(D)} (k,\xi \, \vert \, \Delta_0) & \mbb{S}_{12}^{\,(D)} (k,\xi \, \vert \, \Delta_0) & \mbb{S}_{13}^{\,(D)} (k,\xi \, \vert \, \Delta_0) \\
\mbb{S}_{21}^{\,(D)} (k,\xi \, \vert \, \Delta_0) & \mbb{S}_{22}^{\,(D)} (k,\xi \, \vert \, \Delta_0) & \mbb{S}_{23}^{\,(D)} (k,\xi \, \vert \, \Delta_0)\\
\mbb{S}_{31}^{\,(D)} (k,\xi \, \vert \, \Delta_0) & \mbb{S}_{32}^{\,(D)} (k,\xi \, \vert \, \Delta_0) & \mbb{S}_{33}^{\,(D)} (k,\xi \, \vert \, \Delta_0)
\end{array} \right\}
\end{equation}
for the case of a dice lattice with a bandgap $\Delta_0$ are obtained as

\begin{eqnarray}
\label{dg}
&& \mbb{S}_{11}^{\,(D)} (k,\xi \, \vert \, \Delta_0)/\pi =  \left(\frac{k}{\epsilon_\Delta} \right)^2 \, \delta(\xi) + \left[
1 - \frac{k^2}{2 \epsilon_\Delta^2} + \frac{\Delta_0}{\epsilon_\Delta} \right] \times \delta(\xi - \epsilon_\Delta) +
 \left[
1 - \frac{k^2}{2 \epsilon_\Delta^2} - \frac{\Delta_0}{\epsilon_\Delta} \right] \times \delta(\xi + \epsilon_\Delta)\, , \\
\nonumber
&& \mbb{S}_{12}^{\,(D)} (k,\xi \, \vert \, \Delta_0)/\pi = \frac{k \, \tet{e}^{- i \theta_{\bf k}}}{\sqrt{2}}  \, \left\{
\left[
\frac{1}{\epsilon_\Delta} + \frac{\Delta_0}{2\epsilon_\Delta^2} 
\right] \times \delta(\xi - \epsilon_\Delta) + 
\left[
- \frac{1}{\epsilon_\Delta} + \frac{\Delta_0}{2\epsilon_\Delta^2} 
\right] \times \delta(\xi + \epsilon_\Delta)
 - \frac{\Delta_0}{\epsilon_\Delta^2} \times 
\delta(\xi)
\right\} \, , \\
\nonumber
&& \mbb{S}_{21}^{\,(D)} (k,\xi \, \vert \, \Delta_0)/\pi = \left[\mbb{S}_{21}^{\,(D)} (k,\xi \, \vert \, \Delta_0)\right]^\star/\pi =
\\
\nonumber
&& =\frac{k \, \tet{e}^{i \theta_{\bf k}}}{\sqrt{2}}  \, \left\{
\left[
\frac{1}{\epsilon_\Delta} + \frac{\Delta_0}{2\epsilon_\Delta^2} 
\right] \times \delta(\xi - \epsilon_\Delta) + 
\left[
- \frac{1}{\epsilon_\Delta} + \frac{\Delta_0}{2\epsilon_\Delta^2} 
\right] \times \delta(\xi + \epsilon_\Delta)
 - \frac{\Delta_0}{\epsilon_\Delta^2} \times 
\delta(\xi)
\right\} \, , \\
\nonumber
&& \mbb{S}_{13}^{\,(D)}  (k,\xi \, \vert \, \Delta_0)/\pi = \frac{k^2 \,  \tet{e}^{- 2 i \theta_{\bf k}}}{2 \epsilon_\Delta^2} \times \left[
\delta(\xi - \epsilon_\Delta)  + \delta(\xi + \epsilon_\Delta)  - 2 \delta(\xi) 
\right] \, , \\
\nonumber 
&& \mbb{S}_{31}^{\,(D)}  (k,\xi \, \vert \, \Delta_0)/\pi =
\left[\mbb{S}_{13}^{\,(D)} (k,\xi \, \vert \, \Delta_0)\right]^\star/\pi
=\frac{k^2 \,  \tet{e}^{2 i \theta_{\bf k}}}{2 \epsilon_\Delta^2} \times \left[
\delta(\xi - \epsilon_\Delta)  + \delta(\xi + \epsilon_\Delta)  - 2 \delta(\xi) 
\right] \, , \\
\nonumber 
&& \mbb{S}_{22}^{\,(D)} (k,\xi \, \vert \, \Delta_0)/\pi = 
\left( \frac{\Delta_0}{2 \epsilon_\Delta} \right)^2 \, \delta(\xi) + 
\left[ 1 - \left( \frac{\Delta_0}{2 \epsilon_\Delta} \right)^2 \, \right] \times
\left[  
\delta(\xi - \epsilon_\Delta)  + \delta(\xi + \epsilon_\Delta)
\right]  \, , \\
\nonumber 
&& \mbb{S}_{23}^{\,(D)}  (k,\xi \, \vert \, \Delta_0)/\pi = \frac{k \,  
\tet{e}^{- i \theta_{\bf k}}}{ \sqrt{2} \, \epsilon_\Delta^2} \times 
\left\{\Delta_0 \, \delta(\xi) + \left[\epsilon_\Delta - \frac{\Delta_0}{2}\right] \, \delta(\xi - \epsilon_\Delta)  - \left[\epsilon_\Delta + \frac{\Delta_0}{2} \right] \delta(\xi + \epsilon_\Delta) 
\right\}\, , \\
&& \mbb{S}_{32}^{\,(D)}  (k,\xi \, \vert \, \Delta_0)/\pi = \left[
\mbb{S}_{32}^{\,(D)}  (k,\xi \, \vert \, \Delta_0)/\pi
\right]^\star = \\
\nonumber 
&& \frac{k \,  
\tet{e}^{- i \theta_{\bf k}}}{ \sqrt{2} \, \epsilon_\Delta^2} \times 
\left\{
\Delta_0 \, \delta(\xi) + \left[\epsilon_\Delta - \frac{\Delta_0}{2}\right] \, \delta(\xi - \epsilon_\Delta)  - \left[\epsilon_\Delta + \frac{\Delta_0}{2} \right] \delta(\xi + \epsilon_\Delta) 
\right\} \, , \\
\nonumber 
&& \mbb{S}_{33}^{\,(D)} (k,\xi \, \vert \, \Delta_0)/\pi = \left(\frac{k}{\epsilon_\Delta} \right)^2 \, \delta(\xi) + \left[
1 - \frac{k^2}{2 \epsilon_\Delta^2} - \frac{\Delta_0}{\epsilon_\Delta} \right] \times \delta(\xi - \epsilon_\Delta) +
 \left[
1 - \frac{k^2}{2 \epsilon_\Delta^2} + \frac{\Delta_0}{\epsilon_\Delta} \right] \times \delta(\xi + \epsilon_\Delta)\, .
\end{eqnarray}

\medskip
For a zero badngap, Eq.~\eqref{dg} is reduced to

\begin{eqnarray}
&& \frac{1}{\pi} \,\mbb{S}_{11}^{\,(D)} (k,\xi \, \vert \, \Delta_0=0)  =   \delta(\xi) + \frac{1}{2} [\delta(\xi - k) +
\delta(\xi + k)]\, , \\
\nonumber
&& \frac{1}{\pi} \,\mbb{S}_{12}^{\,(D)} (k,\xi \, \vert \, \Delta_0=0) = \frac{\tet{e}^{- i \theta_{\bf k}}}{\sqrt{2}}  \, \left[
 \delta(\xi - k) - \delta(\xi + k)
\right]
 \, , \\
\nonumber
&& \frac{1}{\pi} \,\mbb{S}_{21}^{\,(D)} (k,\xi \, \vert \, \Delta_0=0) = \frac{1}{\pi} \,\mbb{S}^\star_{12} (k,\xi \, \vert \, \Delta_0=0)
= \frac{\tet{e}^{- i \theta_{\bf k}}}{\sqrt{2}}  \, \left[ \delta(\xi - k) - \delta(\xi + k) \right]
\, , \\
\nonumber 
&& \frac{1}{\pi} \,\mbb{S}_{13}^{\,(D)} (k,\xi \, \vert \, \Delta_0=0) = \frac{1}{2} \, \tet{e}^{- 2 i \theta_{\bf k}} \times \left[
\delta(\xi + k)  + \delta(\xi - k)  - 2 \delta(\xi) 
\right] \, , \\
\nonumber 
&& \frac{1}{\pi} \,\mbb{S}_{31}^{\,(D)} (k,\xi \, \vert \, \Delta_0=0) = \frac{1}{\pi} \,\left[ \mbb{S}_{13}^{\,(D)} (k,\xi \, \vert \, \Delta_0) 
\right]^\star = \frac{1 
}{2} \, \tet{e}^{2 i \theta_{\bf k}} \times \left[ \delta(\xi + k)  + \delta(\xi - k)  - 2 \delta(\xi) 
\right] \, , \\
\nonumber 
&& \frac{1}{\pi} \,\mbb{S}_{22}^{\,(D)} (k,\xi \, \vert \, \Delta_0=0) =  \delta(\xi - k)  + \delta(\xi + k) \, , \\
\nonumber 
&& \frac{1}{\pi} \,\mbb{S}_{23}^{\,(D)} (k,\xi \, \vert \, \Delta_0=0) = \frac{\tet{e}^{- i \theta_{\bf k}}}{ \sqrt{2}} \times 
\left[
\delta(\xi - k) - \delta(\xi + k) 
\right] \, , \\
\nonumber 
&& \frac{1}{\pi} \,\mbb{S}_{32}^{\,(D)} (k,\xi \, \vert \, \Delta_0=0) = \frac{1}{\pi} \,\left[ \mbb{S}_{23}^{\,(D)} (k,\xi \, \vert \, \Delta_0=0)
\right]^\star = \frac{  
\tet{e}^{i \theta_{\bf k}}}{\sqrt{2}} \times 
\left[
\delta(\xi - k) - \delta(\xi + k) 
\right] \, , \\
\nonumber 
&& \frac{1}{\pi} \,\mbb{S}_{33}^{\,(D)} (k,\xi \, \vert \, \Delta_0 = 0) = \frac{1}{\pi} \,\mbb{S}_{11}^{\,(D)} (k,\xi \, \vert \, \Delta_0=0)  =  
\delta(\xi) + \frac{1}{2} [\delta(\xi - k) + \delta(\xi + k)]\, .
\end{eqnarray}

Thus, the sought trace $\mbb{T}(k, \xi \, \vert \, \omega)$ is  

\begin{equation}
\label{traceA}
\mbb{T}(k, \xi \, \vert \, \omega)= \frac{1}{2} \, \int\limits \frac{d^2 {\bf k}}{2 \pi} \, \left[ \mbb{T}_1(k, \xi \, \vert \, \omega) + 
\, \mbb{T}_2 (k,\xi \, \vert \, \omega)
\right] \, , 
\end{equation}

where

\begin{eqnarray}
&& \mbb{T}_1(k , \xi \, \vert \, \omega) = \left[\mbb{S}^\phi_{21} (k,\xi) + \mbb{S}^\phi_{23} (k,\xi) \right] \times
\left[\mbb{S}^\phi_{21} (k,\xi + \omega) + \mbb{S}^\phi_{23} (k,\xi + \omega) \right] + \\
\nonumber
&& + \left[\mbb{S}^\phi_{12} (k,\xi) + \mbb{S}^\phi_{32} (k,\xi) \right] \times
\left[\mbb{S}^\phi_{12} (k,\xi + \omega) + \mbb{S}^\phi_{32} (k,\xi + \omega) \right]
\\
\nonumber 
&& \mbb{T}_2(k , \xi \, \vert \, \omega) =\mbb{S}^\phi_{22} (k,\xi + \omega)  \times \left[\mbb{S}^\phi_{11} (k,\xi) + \mbb{S}^\phi_{33} (k,\xi) \right] + 
\mbb{S}^\phi_{22} (k,\xi)  \times \left[\mbb{S}^\phi_{11} (k,\xi + \omega) + \mbb{S}^\phi_{33} (k,\xi + \omega) \right] 
\end{eqnarray}

\begin{equation}
\label{traceA2}
\mbb{T}^{\,(D)}_{\Delta_0 = 0}(k, \xi \, \vert \, \omega)= \int\limits \frac{d^2 {\bf k}}{4 \pi} \, \left[ \mbb{T}^{\,(D)}_{1,\Delta_0 = 0}(k, \xi \, \vert \, \omega) + \mbb{T}^{\,(D)}_{2,\Delta_0 = 0}(k, \xi \, \vert \, \omega) + 4 \mbb{T}^{\,(D)}_{3,\Delta_0 = 0}(k, \xi \, \vert \, \omega)
\right] \, , 
\end{equation}
where

\begin{eqnarray}
&& \mbb{T}^{\,(D)}_{1,\Delta_0 = 0}(k, \xi \, \vert \, \omega) = \delta(\omega + \xi - k) \times \left[
5 \, \delta(\xi - k ) - 3 \, \delta(\xi +k) \right]
\, , \\ 
\nonumber 
&& \mbb{T}^{\,(D)}_{2,\Delta_0 = 0}(k, \xi \, \vert \, \omega) = \delta(\omega + \xi + k) \times \left[
5 \, \delta(\xi + k ) - 3 \, \delta(\xi - k) \right]
\, , \\ 
\nonumber
&& \mbb{T}^{\,(D)}_{3,\Delta_0 = 0}(k, \xi \, \vert \, \omega) = \delta(\xi) \times \left[
 \delta(\xi +\omega - k) + \delta(\xi +\omega + k)
\right] \, . 
\end{eqnarray}

\begin{equation}
\label{t}
\mbb{T}^{\,(D)}_{\Delta_0 = 0}(k, \xi \, \vert \, \omega)= \int\limits \frac{d^2 {\bf k}}{4 \pi} \, \left[ \mbb{T}^{\,(D)}_{1,\Delta_0 = 0}(k, \xi \, \vert \, \omega) + 4 \mbb{T}^{\,(D)}_{3,\Delta_0 = 0}(k, \xi \, \vert \, \omega)
\right] \, , 
\end{equation}
where

\begin{eqnarray}
&& \mbb{T}^{\,(D)}_{1,\Delta_0 = 0}(k, \xi \, \vert \, \omega) = \sum \limits_{\beta = \pm 1} \delta(\omega + \xi - \beta k) \, 
\times \left[
5 \, \delta(\xi -   \beta k ) - 3 \, \delta(\xi +  \beta k) \right]
\, , \\ 
\nonumber 
&& \mbb{T}^{\,(D)}_{1,\Delta_0 = 0}(k, \xi \, \vert \, \omega) = \delta(\xi) \times \sum \limits_{\beta = \pm 1}
 \delta(\xi +\omega + \beta k) 
 \, . 
\end{eqnarray}

\section{Green's functions and spectral functions for $\alpha-\mc{T}_3$ model}
\label{apc}

\vskip0.5in
The spectral function, generally written by 

\begin{equation} \hat{\mbb{S}}^{\,(\phi)} (k,\xi)= 
\left\{
\begin{array}{ccc}
\mbb{S}^\phi_{11} (k,\xi) & \mbb{S}^\phi_{12} (k,\xi) & \mbb{S}^\phi_{13} (k,\xi) \\
\mbb{S}^\phi_{21} (k,\xi) & \mbb{S}^\phi_{22} (k,\xi) & \mbb{S}^\phi_{23} (k,\xi)\\
\mbb{S}^\phi_{31} (k,\xi) & \mbb{S}^\phi_{32} (k,\xi) & \mbb{S}^\phi_{33} (k,\xi)
\end{array} \right\} \, ,
\end{equation}

has the following elements for an arbitrary $\alpha-\mc{T}_3$ 

\begin{eqnarray}
&& \mbb{S}^{\,(\phi)}_{11} (k,\xi)/\pi = \cos^2 \phi \times \left[ \delta(\xi - k) + \delta(\xi + k)
\right] + 2\, \sin^2 \phi \times \delta(\xi) \, , \\
\nonumber 
&& \mbb{S}^{\,(\phi)}_{12} (k,\xi)/\pi  = \tet{e}^{- i \theta_{\bf k}} \, \cos \phi \times \left[ \delta(\xi - k) - \delta(\xi + k)
\right] \, , \\
\nonumber
&& \mbb{S}^{\,(\phi)}_{21} (k,\xi) = \left[\mbb{S}_{12} (k,\xi) \right]^\star = \pi \, \tet{e}^{i \theta_{\bf k}} \, \cos \phi \times \left[ \delta(\xi - k) - \delta(\xi + k) \right] \, , \\
\nonumber 
&& \mbb{S}^{\,(\phi)}_{13} (k,\xi)/ \pi =   \tet{e}^{- 2 i \theta_{\bf k}} \, \sin (2 \phi) \times \left[ - \frac{1}{2} \, \delta(\xi) + \delta(\xi - k) + \delta(\xi + k) 
\right] \, , \\
\nonumber
&& \mbb{S}^{\,(\phi)}_{31} (k,\xi) = \left[\mbb{S}_{13}^{\,(\phi)} (k,\xi) \right]^\star = \pi \, \tet{e}^{- 2 i \theta_{\bf k}} \, \sin (2 \phi) \times \left[ \delta(\xi - k) + \delta(\xi + k) - \frac{1}{2} \, \delta(\xi) \right] \, , \\
\nonumber
&& \mbb{S}^{\,(\phi)}_{22} (k,\xi)/ \pi =  \delta(\xi - k) + \delta(\xi + k) \, , \\
\nonumber
&& \mbb{S}^{\,(\phi)}_{23} (k,\xi) = \pi \, \tet{e}^{- i \theta_{\bf k}} \, \sin \phi \times \left[ \delta(\xi - k) - \delta(\xi + k)
\right] \, , \\
\nonumber
&& \mbb{S}^{\,(\phi)}_{32} (k,\xi) = \mbb{S}^\star_{23} (k,\xi) = \pi \, \tet{e}^{i \theta_{\bf k}} \, \sin \phi \times \left[ \delta(\xi - k) - \delta(\xi + k)
\right] \, , \\
\nonumber
&& \mbb{S}^{\,(\phi)}_{33} (k,\xi) =  \sin^2 \phi \times \left[ \delta(\xi - k) + \delta(\xi + k)
\right] + 2\, \cos^2 \phi \times \delta(\xi) \, .
\end{eqnarray}

Thus, the trace in Eq.\eqref{tr} is obtained as 

\begin{equation}
\label{traceAA}
\mbb{T}^{[\phi]}(\xi) = \mbb{T}_1(k , \xi \, \vert \, \omega) \, \cos^2 \phi + \frac{1}{2} \, \mbb{T}_2 (k , \xi \, \vert \, \omega) \, \sin (2 \phi) + \mbb{T}_3 k , (\xi \, \vert \, \omega) \, \sin^2 \phi 
\end{equation}

where

\begin{eqnarray}
&& \mbb{T}_1(k , \xi \, \vert \, \omega) = \mbb{S}^\phi_{11} (k,\xi + \omega) \, \mbb{S}^\phi_{22} (k,\xi) + \mbb{S}^\phi_{11} (k,\xi) \, \mbb{S}^\phi_{22} (k,\xi + \omega) \\
\nonumber 
&& \mbb{T}_2(k , \xi \, \vert \, \omega) = \sum \limits_{\nu = \pm 1} \left\{ \mbb{S}^\phi_{21} \left[ k, \xi + (\nu + 1)\, \frac{\omega}{2} \right] \, \mbb{S}^\phi_{23} \left[ k, \xi + (\nu + 1)\, \frac{\omega}{2}\right] + \left( 2 \longleftrightarrow 1 \, \&  \, 3 \longleftrightarrow 2 \right) \right\} + \\
\nonumber 
&& + \sum \limits_{\nu = \pm 1} \mbb{S}^\phi_{22} \left[ k, \xi + (\nu + 1)\, \frac{\omega}{2} \right] \left\{
\mbb{S}^\phi_{13} \left[ k, \xi + (\nu + 1)\, \frac{\omega}{2} \right] + \mbb{S}^\phi_{31} \left[ k, \xi + (\nu + 1)\, \frac{\omega}{2} \right]
\right\}
\\
\nonumber 
&& \mbb{T}_3(k , \xi \, \vert \, \omega) = \mbb{S}^\phi_{22} (k,\xi + \omega) \, \mbb{S}^\phi_{33} (k,\xi) + \mbb{S}^\phi_{22} (k,\xi) \, \mbb{S}^\phi_{33} (k,\xi + \omega) \, .
 \end{eqnarray}

Here, we preliminary excluded terms with a angular dependence  $\backsim \tet{e}^{\pm 2 i \theta_{\bf k}}$, $\backsim \tet{e}^{\pm 3 i \theta_{\bf k}}$ and $\backsim \tet{e}^{\pm 4 i \theta_{\bf k}}$, which would become zero after we perform the angular integration. It is easy to see that the last term of $\mbb{T}_2(k , \xi \, \vert \, \omega)$ equals to zero for all $\phi \neq \pi/4$ (except for a dice lattice). 

\medskip 

For $\phi = 0$, we obviously recover a well-known graphene limit: 

\begin{eqnarray}
\label{traceF}
&& \mbb{T}^{\,(\phi)}(\xi) = \mbb{S}^\phi_{11} (k,\xi + \omega) \, \mbb{S}^\phi_{22} (k,\xi) + \mbb{S}^\phi_{11} (k,\xi) \, \mbb{S}^\phi_{22} (k,\xi + \omega) = \\
\nonumber  
&& = 2 \pi^2 \left[ \delta(\xi - k) + \delta(\xi + k) \right] \times \left[ \delta(\xi - k + \omega) + \delta(\xi + k + \omega) \right] \, .
\end{eqnarray}

while the other limit $\phi= \pi/4$ is a gapless dice lattice discussed in the previous Appendix \ref{apb}.

\bibliography{OBib}
\end{document}